\documentclass[12pt]{article}

\begin{document}

\begin{center}
{\large \bf Analog of Electroweak Model for Contracted Gauge Group}
\end{center}

\begin{center}
N.A.~Gromov \\
Department of Mahematics, Komi Science Center UrD, RAS, \\
Kommunisticheskaya st. 24, Syktyvkar 167982, Russia \\
E-mail: gromov@dm.komisc.ru
\end{center}

\begin{abstract}
The  limiting case of the bosonic part of the standard Electroweak Model, which correspond to the contracted gauge group $SU(2;j)\times U(1)$ is discussed.  The Higgs boson, Z-boson and electromagnetic fields become an external ones with respect to the W-bosons fields. The W-bosons fields do not effect on these external fields. The masses of the all  particles of the Electroweak Model remain the same under contraction.
\end{abstract}


PACS number:  12.15-y

\section{Introduction }

The Standard Electroweak Model  based on gauge group $ SU(2)\times U(1)$ gives a good  description of electroweak processes. One of its important ingredient is the simple group $ SU(2).$ On the other hand, in physics it is well known group contraction \cite{IW-53}, which transforms, for example, a simple or semisimple group to a non-semisimple one.
For better understanding of a physical system it is useful to  investigate  its limits for limiting values of their physical parameters. 

In the present paper we construct the analog of the bosonic part of the standard Electroweak Model as the gauge theory with the contracted non-semisimple group $SU(2;j)\times U(1)$. 
The gauge theories for non-semisimple groups which Lie algebras admit invariant non-degenerate metrics was considered in \cite{NW-93}, \cite{T-95}. The  construction given in   \cite{S-07} is based on an observation: the underlying group of the standard Electroweak Model  can be represented as a semidirect product of $U(1)$ and $SU(2).$

Most often the fundamental representation of the contracted special unitary group $SU(2;j)$ is represented by the triangular matrix
$$
u=\left(\begin{array}{cc}
a & 0 \\
b	 & \bar{a}
\end{array} \right),\quad \det u=a\bar{a}=1
$$
with complex matrix elements $a,b\in{\bf C}$. But there is another possibility \cite{Gr-94}.  
Instead of matrix with complex  elements we can use matrix with nilpotent elements in the form
$$ 
u(\iota)= 
\left(\begin{array}{cc}
\alpha	 & \iota\beta   \\
-\iota\bar{\beta}	 & \bar{\alpha}
\end{array} \right), \quad  \det u(\iota)=|\alpha|^2=1,  \quad
\alpha, \beta \in {\bf C}, 
$$
where nilpotent unit $\iota\neq 0,$ but $\iota^2=0.$ These two possibilities correspond to the nonsymmetric appearance
of contraction parameter $j$ in the first case, when the matrix element $u_{12}$ is multiplied by $j^2=\iota^2=0$ and the symmetric one, when matrix elements $u_{12}, u_{21}$  are multiplied by $j=\iota.$

In Sec. II, we recall the definition and properties of the contracted group $SU(2;j)$. In Sec. III, we step by step modify
the main points of the Electroweak Model for the gauge group $SU(2;j)\times U(1)$.
The use of the representation with the symmetric arranged contraction parameter enables us to find transformation properties (\ref{g14}) of gauge fields. After that the Lagragian of the contracted model can be very easy obtained from the standard one.   

Two interpretations of the contracted model are discussed. In the first one 
the Higgs boson, Z-boson and electromagnetic fields  can be regarded as an external fields with respect to the W-bosons fields and the last ones do not effect on these external fields. The masses of the all  particles of the Electroweak Model do not changed  under contraction.

\section{Contracted Special Unitary group  $SU(2;j)$ }

Let us regard  two dimensional complex fibered vector space $\Phi_2(j)$ with one dimensional base $\left\{\phi_2\right\}$
and one dimensional fiber $\left\{\phi_1\right\}$ \cite{Gr-94}. This space has two hermitian forms: first  in the base 
$\bar{\phi_2}\phi_2=|\phi_2|^2$ and second in the fiber $\bar{\phi_1}\phi_1=|\phi_1|^2,$ where bar denotes complex conjugation. Both forms can be written by one formula
\begin{equation}
\phi^\dagger\phi(j)=j^2|\phi_1|^2+|\phi_2|^2,
\label{g1}
\end{equation}  
where $\phi^\dagger=(j\bar{\phi_1},\bar{\phi_2}), $
parameter $j=1, \iota$ and $\iota$ is nilpotent unit $\iota^2=0.$   
We shall demand that the following heuristic rules be fulfiled: for a real or complex $a$ the expression $\frac{a}{\iota}$
is defined only for $a=0,$ however $\frac{\iota}{\iota}=1.$
The usual complex vector  space $\Phi_2$ is obtained for $j=1.$
Inverse correspondence is given by
\begin{equation}
|\phi_2|^2=\phi^\dagger\phi(j)_{|j=\iota}; \quad
|\phi_1|^2=\frac{1}{j^2}\phi^\dagger\phi(j)_{|j=\iota}, \; \mbox{for}\;\; \phi_2=0.
\label{g2}
\end{equation}  

 The special unitary group $SU(2;j)$ is defined as a transformation group of $\Phi_2(j)$ which keep invariant the hermitian form (\ref{g1}), i.e.
$$ 
\phi'(j)=
\left(\begin{array}{c}
j\phi'_1 \\
\phi'_2
\end{array} \right)
=\left(\begin{array}{cc}
	\alpha & j\beta   \\
-j\bar{\beta}	 & \bar{\alpha}
\end{array} \right)
\left(\begin{array}{c}
j\phi_1 \\
\phi_2
\end{array} \right)
=u(j)\phi(j), \quad
$$
\begin{equation}
\det u(j)=|\alpha|^2+j^2|\beta|^2=1.
\label{g3}
\end{equation}  

 The fundumental representation of the one-parameter subgroups of  $SU(2;j)$  are easily obtained
\begin{equation}
\omega_1(\alpha_1;j)=e^{i\alpha_1T_1(j)}=\left(\begin{array}{cc}
	\cos \frac{j\alpha_1}{2} & i\sin \frac{j\alpha_1}{2} \\
i\sin \frac{j\alpha_1}{2}	 & \cos \frac{j\alpha_1}{2}
\end{array} \right),
\label{g4}
\end{equation}  
\begin{equation}
\omega_2(\alpha_2;j)=e^{i\alpha_2T_2(j)}=\left(\begin{array}{cc}
	\cos \frac{j\alpha_2}{2} & \sin \frac{j\alpha_2}{2} \\
-\sin \frac{j\alpha_2}{2}	 & \cos \frac{j\alpha_2}{2}
\end{array} \right),
\label{g5}
\end{equation}  
\begin{equation}
\omega_3(\alpha_3;j)=e^{i\alpha_3T_3(j)}=\left(\begin{array}{cc}
	e^{i\frac{\alpha_3}{2}} & 0 \\
0	 & e^{-i\frac{\alpha_3}{2}}
\end{array} \right).
\label{g6}
\end{equation}  
The corresponding generators 
\begin{equation}    
  T_1(j)= \frac{j}{2}\left(\begin{array}{cc}
	0 & 1 \\
	1 & 0
\end{array} \right), \quad 
T_2(j)= \frac{j}{2}\left(\begin{array}{cc}
	0 & -i \\
	i & 0
\end{array} \right), \quad 
T_3(j)= \frac{1}{2}\left(\begin{array}{cc}
	1 & 0 \\
	0 & -1
\end{array} \right) 
\label{g7}
\end{equation} 
with commutation relations
$$  
[T_1(j),T_2(j)]=-ij^2T_3(j), \quad [T_3(j),T_1(j)]=-iT_2(j), 
$$
\begin{equation} 
 [T_2(j),T_3(j)]=-iT_1(j),
\label{g8}
\end{equation}
 form the Lie algebra $su(2;j).$
 
There are two more or less equivalent way of group contraction. We can put the contraction parameter equal to the nilpotent unit $j=\iota$  or  tend it to zero $j\rightarrow 0$.  Sometimes it is  convenient to use mathematical approach, sometimes physical one. For example the matrix $u(j)$ in (\ref{g3}) has nilpotent non-diagonal elements for $j=\iota$,
whereas for $j\rightarrow 0$ they are formally equal to zero. Nevertheless both approaches lead to the same final results.
 
Let us describe the contracted group $SU(2;\iota)$ in detail. For $j=\iota$ it follows from (\ref{g3}) that 
$\det u(\iota)=|\alpha|^2=1,$ i.e. $\alpha=e^{i\varphi},$ therefore
\begin{equation} 
u(\iota)= 
\left(\begin{array}{cc}
e^{i\varphi}	 & \iota\beta   \\
-\iota\bar{\beta}	 & e^{-i\varphi}
\end{array} \right), \quad
\beta=\beta_1+i\beta_2 \in {\bf C}. 
\label{g9}
\end{equation}
Functions of nilpotent arguments are defined by their Taylor expansion, in particular, 
$\cos\iota x=1,\;\sin\iota x=\iota x.$ Then one-parameter subgroups of  $SU(2;\iota)$ take the form 
$$  
\omega_1(\alpha_1;\iota)=e^{i\alpha_1T_1(\iota)}=\left(\begin{array}{cc}
	1 & \iota i\frac{\alpha_1}{2} \\
	\iota i\frac{\alpha_1}{2} & 1
\end{array} \right), 
$$
\begin{equation}
\omega_2(\alpha_2;\iota)=e^{i\alpha_2T_2(\iota)}=\left(\begin{array}{cc}
1 & \iota \frac{\alpha_2}{2} \\
-\iota \frac{\alpha_2}{2}	 & 1
\end{array} \right).
\label{g11}
\end{equation}  
The third subgroup does not changed and is given by (\ref{g6}).
The simple group $SU(2)$ is contracted to the non-semisimple  group $SU(2;\iota)$, which is isomorphic to the real  Euclid group $E(2).$
First two generators of the Lie algebra $su(2;\iota)$ are commute $[T_1(\iota),T_2(\iota)]=0$ and the rest commutators are given by (\ref{g8}). For the general element $T(\iota)=a_1T_1(\iota)+a_2T_2(\iota)+a_3T_3(\iota)$ of $su(2;\iota)$ the corresponding group element of $SU(2;\iota)$ is as follows 
\begin{equation} 
u(\iota)=e^{iT(\iota)}=
\left(\begin{array}{cc}
e^{i\frac{a_3}{2}}	 & \iota\frac{a}{a_3}\sin \frac{a_3}{2}   \\
-\iota\frac{\bar{a}}{a_3}\sin \frac{a_3}{2} & e^{-i\frac{a_3}{2}}
\end{array} \right), \quad a=a_2+ia_1.
\label{g12}
\end{equation}

The actions of the unitary group $U(1)$ and the electromagnetic subgroup $U(1)_{em}$ 
in the   fibered  space $\Phi_2(\iota)$ are given by the same matrices as on the space $\Phi_2$, namely
\begin{equation}
\omega(\beta)=e^{i\beta Y}=\left(\begin{array}{cc}
	e^{i\frac{\beta}{2}} & 0 \\
0	 & e^{i\frac{\beta}{2}}
\end{array} \right), \quad
\omega_{em}(\gamma)=e^{i\gamma Q}=\left(\begin{array}{cc}
	1+i\gamma & 0 \\
0	 & 1
\end{array} \right).
\label{g13}
\end{equation}  

Representations of groups $SU(2;\iota), U(1), U(1)_{em}$   are linear ones,  that is they are realised  by linear operators in  the   fibered  space $\Phi_2(\iota)$.

\section{ Electroweak Model for $SU(2;j)\times U(1)$ gauge group}

 We shall follow the book \cite{R-99} in description of standard Electroweak Model.  
The bosonic sector of Electroweak Model 
is $SU(2;j)\times U(1)$ gauge theory in the space $\Phi_2(j) $  of fundamental representation of $SU(2;j).$ The  bosonic Lagrangian is given by the sum
\begin{equation}
L_B(j)=L_A(j) + L_{\phi}(j),
\label{eq1}
\end{equation}
where
$$  
 L_A(j)=\frac{1}{8g^2}\mbox{Tr}(F_{\mu\nu}(j))^2-\frac{1}{4}(B_{\mu\nu})^2= 
 $$
 \begin{equation}
=  -\frac{1}{4}[j^2(F_{\mu\nu}^1)^2+j^2(F_{\mu\nu}^2)^2+(F_{\mu\nu}^3)^2]-\frac{1}{4}(B_{\mu\nu})^2
\label{eq2}
\end{equation}
is the gauge field Lagrangian for $SU(2;j)\times U(1)$ group and
\begin{equation}   
  L_{\phi}(j)= \frac{1}{2}(D_\mu \phi(j))^{\dagger}D_\mu \phi(j) -V(\phi(j))
\label{eq3}
\end{equation}  
is the matter field Lagrangian (summation on the repeating Greek indexes is always understood). Here 
$ \phi(j)= \left(
\begin{array}{c}
	j\phi_1 \\
	\phi_2
\end{array} \right) \in \Phi_2(j),\;$
  $D_{\mu}$ are the covariant derivatives
 \begin{equation}
D_\mu\phi(j)=\partial_\mu\phi(j) -ig\left(\sum_{k=1}^{3}T_k(j)A^k_\mu \right)\phi(j)-ig'YB_\mu\phi(j),
\label{eq4}
\end{equation} 
where $T_k(j)$ are given by (\ref{g7})
 and 
$Y=\frac{1}{2}{\bf 1}$ is generator of $U(1).$ 
The gauge fields 
$$ 
A_\mu (x;j)=-ig\sum_{k=1}^{3}T_k(j)A^k_\mu (x),\quad B_\mu (x)=-ig'B_\mu (x)
$$
 take their values in Lie algebras $su(2;j),$  $u(1)$ respectively,  and the stress tensors are
$$ 
F_{\mu\nu}(x;j)={\cal F}_{\mu\nu}(x;j)+[A_\mu(x;j),A_\nu(x;j)],\quad  B_{\mu\nu}=\partial_{\mu}B_{\nu}-\partial_{\nu}B_{\mu}.         
$$
The potential $V(\phi(j))$ in (\ref{eq3}) is introduced by hand in a special form (sombrero)  
\begin{equation}   
  V(\phi(j))=\frac{\lambda }{4}\left(\phi^{\dagger}(j)\phi(j)- v^2\right)^2, 
\label{eq5}
\end{equation}  
where $\lambda, v $ are constants. 

The Lagrangian   $L_B(j)$ (\ref{eq1}) describe  massless fields. To generate   mass terms for the vector bosons without breaking the gauge invariance one uses the Higgs mechanism. One of  $L_B(j)$ 
ground states
$$   
  \phi^{vac}=\left(\begin{array}{c}
	0  \\
	v 
\end{array} \right), \quad  A_\mu^k=B_\mu=0
$$
is taken as a vacuum state of the model, and small field excitations 
$$   
 \phi_1(x), \quad \phi_2(x)=v+\chi(x), \quad A_\mu^a(x), \quad B_\mu(x)
$$
with respect to the vacuum  are regarded.
 The matrix  
  $
  Q=Y+T^3=\left(\begin{array}{cc}
	1 & 0 \\
	0 & 0
\end{array} \right),
  $
  which annihilates the ground state $Q\phi^{vac}=0,$ is the generator of the electromagnetic subgroup
    $U(1)_{em}. $ 
The new fields 
 $$ 
  {W_\mu^{\pm}=\frac{1}{\sqrt{2}}\left(A_\mu^1\mp iA_\mu^2  \right)}, \quad
   { Z_\mu =\frac{1}{\sqrt{g^2+g'^2}}\left( gA_\mu^3-g'B_\mu \right)},
 $$
 \begin{equation}
 { A_\mu =\frac{1}{\sqrt{g^2+g'^2}}\left( g'A_\mu^3+gB_\mu \right)} 
  \label{eq5-1}
\end{equation}   
are introduced, where  $W_\mu^{\pm} $ are complex  $\bar{W}_\mu^{-}=W_\mu^{+} $ and $Z_\mu, A_\mu $ are real. 

For small fields the Lagrangian $L_B$ (\ref{eq1}) can be rewritten in the form
$$
L_B(j)=L_0(j) + L_{int}(j),
$$
where $L_0(j)=L_B^{(2)}(j)$ is the usual second order Lagrangian  for free vector and scalar bosons,   
and higher order terms $L_{int}(j)$ are regarded as field interactions. 
   The second order terms of the Lagrangian (\ref{eq1}) are as follows 
     \begin{eqnarray}
L_B^{(2)}(j)=   \frac{1}{2}\left(\partial_\mu\chi \right)^2 -\frac{1}{2}m_{\chi}^2\chi^2
- {\frac{1}{4}{\cal Z}_{\mu\nu}{\cal Z}_{\mu\nu}+\frac{1}{2}m_Z^2Z_\mu Z_\mu} 
-{\frac{1}{4}{\cal F}_{\mu\nu}{\cal F}_{\mu\nu}}  \nonumber\\
  + j^2\left\{ -{\frac{1}{2}{\cal W}_{\mu\nu}^{+}{\cal W}_{\mu\nu}^{-}+m_W^2W_\mu^{+}W_\mu^{-} }\right\}
  = L^{(2)}_1+ j^2L_W^{(2)},  
 \label{eq6}
\end{eqnarray} 
where 
$
{\cal Z}_{\mu\nu}=\partial_\mu Z_\nu-\partial_\nu Z_\mu, \; 
{\cal F}_{\mu\nu}=\partial_\mu A_\nu-\partial_\nu A_\mu, \; 
{\cal W^{\pm}}_{\mu\nu}=\partial_\mu W^{\pm}_\nu-\partial_\nu W^{\pm}_\mu. \; 
$  
 
For $j=1$ Lagrangian (\ref{eq6}) describe massive vector fields  {$W_\mu^{\pm}$ with identical mass  $m_W=\frac{1}{2}gv$} (charded $W$-bosons), massless  vector field {$A_\mu, \; m_{A}=0$} (photon),
massive vector field    {$Z_\mu,$ with the mass  $m_Z=\frac{v}{2}\sqrt{g^2+g'^2}$} (neutral $Z$-boson)
and massive scalar field {$ \chi,\; m_{\chi}=\sqrt{2\lambda}v$} (Higgs boson).
  $W$- and $Z$-bosons have been    observed and have the masses 
  $  m_W=80 GeV,$ $ m_Z=91 GeV.  $
 Higgs boson has not been  experimentally verified up to now.


The fibered space $\Phi_2(j)$ can be obtained from $\Phi_2$ by substituting $j\phi_1$ instead of $\phi_1.$
Substitution $\phi_1 \rightarrow j\phi_1$ induces another ones for Lie algebra generators
$T_1 \rightarrow jT_1,\; T_2 \rightarrow jT_2,\;T_3 \rightarrow T_3. $
As far as the gauge fields take their values in Lie algebra, we can substitute gauge fields instead of transformation of generators, namely
\begin{equation}
A_{\mu}^1 \rightarrow jA_{\mu}^1, \;\; A_{\mu}^2 \rightarrow jA_{\mu}^2,\; \;A_{\mu}^3 \rightarrow A_{\mu}^3, \;\;
B_{\mu} \rightarrow B_{\mu}.
\label{g14}
\end{equation}  
For the  gauge fields (\ref{eq5-1}) and the corresponding stress tensors these substitutions are as follows 
$$
W_{\mu}^{\pm} \rightarrow jW_{\mu}^{\pm}, \;\; Z_{\mu} \rightarrow Z_{\mu},\; \;A_{\mu} \rightarrow A_{\mu},
$$
\begin{equation}
{\cal W}^{+}_{\mu\nu} \rightarrow j{\cal W}^{+}_{\mu\nu},\;\;{\cal Z}_{\mu\nu} \rightarrow {\cal Z}_{\mu\nu},\;\;
{\cal F}_{\mu\nu} \rightarrow {\cal F}_{\mu\nu}.
\label{g15}
\end{equation} 
The matter field $\phi_2$ does not transformed as well as its small part $\chi$. 
These substitutions in the standard Lagrangian $L_B^{(2)}$ of Electroweak Model give rise to (\ref{eq6}).

Let contraction parameter tends to zero $j^2\rightarrow 0$, then the contribution of W-bosons fields to the Lagrangian (\ref{eq6}) will be small in comparison  with Higgs boson, Z-boson and electromagnetic fields. In other words the limit model includes only three last fields and charded W-bosons fields does not effect on these fields.
The  part $L_W^{(2)}$ form a new Lagrangian for W-bosons fields. The appearance of two Lagrangians is in correspondence with two hermitian forms of fibered  space $\Phi_2(\iota)$, which are invariant under the action of  contracted gauge group $SU(2;\iota)$. Higgs boson, Z-boson and electromagnetic fields are appeared as external fields with respect to
the W-bosons fields. 

In mathematical language the  field space $\left\{\chi, A_{\mu}, Z_{\mu}, W_{\mu}^{\pm}\right\}$
is fibered after contraction to the base $\left\{\chi, A_{\mu}, Z_{\mu}\right\}$ and the fiber 
$\left\{W_{\mu}^{\pm}\right\}.$ 
(In order to avoid terminological misunderstanding let us stress that we  have in view locally trivial fibering, which
is defined by the projection $pr:\; \left\{\chi, A_{\mu}, Z_{\mu}, W_{\mu}^{\pm}\right\} \rightarrow \left\{\chi, A_{\mu}, Z_{\mu}\right\}$ in the field space. 
This fibering has nothing to do with the principal bundle.)
Then $L^{(2)}_1$ in (\ref{eq6}) presents Lagrangian in the base and $L_W^{(2)}$ is Lagrangian in the fiber.  In general, properties of a fiber are dependent on a points of a base and not the reverse.
In this sense fields in the base are external with respect to fields in the fiber. 

To demonstrate this statement let us transform the formula (6.59) in \cite{R-99} for interactions of the electromagnetic field with all other fields by substitution (\ref{g15})
\begin{equation}
L^{em}_{int}= -\frac{1}{4}{\cal H}_{\mu\nu}^2 - j^2\frac{1}{2}\left\{|{\cal W}_{\mu\nu}^{-}|^2 + 
ig{\cal H}_{\mu\nu}{\cal P}_{\mu\nu} +j^2\frac{g^2}{2}{\cal P}_{\mu\nu}^2  \right\},
\label{g17}
\end{equation}
where
\begin{equation}
{\cal H}_{\mu\nu}={\cal F}_{\mu\nu}\sin\theta +  {\cal Z}_{\mu\nu}\cos\theta, \quad
{\cal P}_{\mu\nu}={\cal W}_{\mu}^{-}{\cal W}_{\nu}^{+} - {\cal W}_{\nu}^{-}{\cal W}_{\mu}^{+}.  
\label{g18}
\end{equation}
The first term in (\ref{g17})
\begin{equation}
-\frac{1}{4}{\cal H}_{\mu\nu}^2=-\frac{1}{4}\left\{{\cal F}_{\mu\nu}^2\sin^2\theta + 
{\cal F}_{\mu\nu}{\cal Z}_{\mu\nu}\sin2\theta + {\cal Z}_{\mu\nu}^2\cos^2\theta  \right\}
\label{g19}
\end{equation}
describes self-action of the electromagnetic field, interaction with Z-boson field and self-action of Z-boson field (all in the base). 
The last term in (\ref{g17}) disappears as having fourth order in $j\rightarrow 0$ and we obtain the following Lagrangian in the fiber
\begin{equation}
L_f=-{\frac{1}{2}{\cal W}_{\mu\nu}^{+}{\cal W}_{\mu\nu}^{-}+m_W^2W_\mu^{+}W_\mu^{-} }
- \frac{1}{2}\left\{|{\cal W}_{\mu\nu}^{-}|^2 + 
ig{\cal H}_{\mu\nu}{\cal P}_{\mu\nu}   \right\},
\label{g20}
\end{equation}
which describes self-actions and interactions of W-bosons fields as well as their interactions with external electromagnetic and Z-boson fields. Let us note that interactions in contracted model are more simple as compared with the standard Electroweak Model due to nullification of some terms.

\section{Conclusion}

The suggested  contracted model  with the gauge group $SU(2;j)\times U(1)$, where $j=\iota$ or $j \rightarrow 0$,
corresponds to the limiting case of the standard Electroweak Model. The masses of the all  particles involved in the Electroweak Model remain the same under contraction, but interactions of the fields are changed in two aspects. 
Firstly  all field interactions become simpler.   
Secondly  interrelations  of the fields become more complicated. All fields are divided on two class: fields in the base
(Higgs boson, Z-boson and electromagnetic fields) and fields in the fiber (W-bosons fields). 
The fields of the first class are external with respect to the last one. The Higgs boson, Z-boson and electromagnetic fields can interact with W-bosons fields, but in the limit W-bosons fields do not effect on these fields.

The author is grateful to V.V. Kuratov for helpful discussions.
This work has been supported in part by the Russian Foundation for Basic Research, grant 08-01-90010-Bel-a.


\end{document}